%% file: main_acm.tex
\documentclass[format=sigconf, review=false, anonymous=false, screen, dvipsnames]{lib-acm/acmart}

\setcopyright{acmlicensed}
\copyrightyear{2018}
\acmYear{2018}
\acmDOI{XXXXXXX.XXXXXXX}

\acmConference[]{}{}{}
\acmYear{}
\copyrightyear{}
\acmPrice{}
\acmDOI{}
\acmISBN{}
\setcopyright{none}

\usepackage{booktabs} 

\input{_setup.tex} 

\begin{document}

\tolerance=400 

%
\title[What is (H)CI: Why Does the ``Human'' Matter?]{What is (H)CI: Why Does the ``Human'' Matter? }


\author{Sejal Agarwal}
\orcid{0009-0005-2108-4180}
\authornote{All authors contributed equally to this special session proposal.}
\affiliation{%
  \institution{Cheriton School of Computer Science, University of Waterloo}
  \country{Waterloo, Ontario, Canada}
}
\email{s97agarw@uwaterloo.ca}

\author{Delara Forghani}
\orcid{0009-0004-2710-5694}
\authornotemark[1]
\affiliation{%
  \institution{Cheriton School of Computer Science, University of Waterloo}
  \country{Waterloo, Ontario, Canada}
}
\email{delara.forghani@uwaterloo.ca}

\author{Brandon Lit}
\orcid{0009-0002-9418-2695}
\authornotemark[1]
\affiliation{%
  \institution{Cheriton School of Computer Science, University of Waterloo}
  \country{Waterloo, Ontario, Canada}
}
\email{bjlit@uwaterloo.ca}

\author{Thomas Driscoll}
\orcid{0009-0002-7661-8811}
\authornotemark[1]
\affiliation{%
  \institution{Cheriton School of Computer Science, University of Waterloo}
  \country{Waterloo, Ontario, Canada}
}
\email{thomas.driscoll@uwaterloo.ca}

\author{Anthony Maocheia-Ricci}
\orcid{0009-0002-3881-7166}
\authornotemark[1]
\affiliation{%
  \institution{Cheriton School of Computer Science, University of Waterloo}
  \country{Waterloo, Ontario, Canada}
}
\email{anthony.maocheia-ricci@uwaterloo.ca}

\renewcommand{\shortauthors}{Agarwal et al.}

\begin{abstract}
\input{_abstract}

\end{abstract}

%
%




\maketitle

\input{_body.tex}



\bibliographystyle{lib-acm/ACM-Reference-Format}
\bibliography{_references.bib}





\appendix
\makeatother
\clearpage
\renewcommand\thefigure{\thesection.\arabic{figure}}
\renewcommand\thetable{\thesection.\arabic{table}}
\setcounter{figure}{0}
\setcounter{table}{0}

\end{document}

%% file: _setup.tex
\usepackage{exii-macros}

\usepackage{booktabs}

\showrevisions{REVISIONGREEN}

\makeatletter
\g@addto@macro\normalsize{%
  \setlength\abovedisplayshortskip{-9pt}
  \setlength\belowdisplayshortskip{3pt}
}
\makeatother



%% file: _abstract.tex

Human-Computer Interaction (HCI) is a diverse field bringing together theories and methods from fields such as computer science, psychology, and human factors. Historically, HCI has focused on the human through ``user'' or ``human'' centered design, where the focus was either on information processing or understanding people and their concerns with respect to technology. However, amid the increasing adoption of generative AI tools, this workshop explores two critical questions in regards to HCI: \textit{What is HCI?} and \textit{Why does the ``human'' matter?} We aim to bring together researchers from diverse disciplines to reflect on these questions. Through guided discussions, group brainstorming, and reflection, we explore what HCI means, what the field may look like in the future, and why it is important to remember the ``human'' aspect of the field.


%% file: _body.tex


\input{1-motivation}

\input{2-organizers}

\input{3-plans}

\input{4-length}

\input{5-cfp}

%% file: 1-motivation.tex
\section{Motivation}




Human-Computer Interaction (HCI) is both a multidisciplinary and an interdisciplinary field of study that brings together theories and methods from many fields, including cognitive psychology, computer science, and human factors \cite{figma_what_is_hci}. HCI itself is also an extremely diverse field that encompasses a wide-range of sub-fields, including but not limited to: usable security \cite{Spagnolli_GetAwayWithCyberAttacks_2022, Gutfleisch_UsableSecurity_2022}, extended reality \cite{Dogan_AugmentedIntelliXR_2024}, social and civic computing \cite{Gordon_JuryLearning_2022, Jasim_CommunityClick_2021}, creativity support tools \cite{Shneiderman_CreatingCreativity_2000}, and human performance with input devices \cite{Chen_Curves_Ahead_2025}. 

With respect to the multi- and interdisciplinary nature of HCI, the motivation behind this workshop is to bring together researchers from diverse disciplines to reflect on and discuss various factors that influence and impact the field. In particular, we seek to discuss personal definitions and perceptions of what HCI is, its impact on other disciplines, the various contribution types in HCI (such as novel systems and interaction techniques, user perception studies), and the importance of human-centred study. With the continued and increased adoption of generative AI (GenAI) tools, we also want to explore what the future of HCI may look like and reflect on a core aspect of this field: \textit{Why does the ``human'' matter?}

\subsection{History of HCI}
HCI has evolved from a hardware-centred, expert-driven activity into a ubiquitous, human-centred partnership~\cite{grudin2017}. In its earliest phase, humans supported early computers by manually operating and maintaining them. With the advent of transistors in the mid-20th century, computers became collaborative tools, introducing ideas like human–computer symbiosis, interactive graphics, and systems to augment human intellect~\cite{grudin2017}. As computing hardware became smaller and more accessible from the 1960s to the 1990s, HCI shifted toward discretionary use, where non-experts chose to engage with personal computers through user-friendly innovations like graphical user interfaces. The rise of the internet expanded HCI by emphasizing user experience and user behaviour~\cite{grudin2017}. 

HCI research practices can be found in computer graphics, operating systems, human factors, ergonomics, industrial engineering, cognitive psychology, and systems-oriented areas of computer science~\cite{hewett1992,jacko2012}. HCI in computer graphics involves designing graphical interfaces for humans to interact with, while in operating systems, it involves developing mechanisms for input/output devices. Human factors and ergonomics, together with cognitive science, have led to the development of “cognitive ergonomics” and “cognitive engineering.” Also, in cognitive psychology, it examines how humans learn and mentally represent systems~\cite{hewett1992}.

Computer–Human Interaction (CHI) represents a narrower area within HCI~\cite{jacko2012}. It is mainly linked to computer science and is closely associated with the Association for Computing Machinery Special Interest Group on Computer–Human Interaction (ACM SIGCHI) and its annual CHI Conference~\cite{jacko2012,grudin2017}. According to ACM SIGCHI, HCI is a discipline concerned with the design, evaluation and implementation of interactive computing systems for human use and with the study of major phenomena surrounding them~\cite{hewett1992}. 

Despite the diverse background that HCI has drawn from, the largest conferences and research areas for HCI work has firmly been rooted within computer science \cite{jacko2012}. While this appears to be a logical decision, applying the concept of HCI has historically been challenging. In an ever growing digital world HCI is implicitly everywhere, with implications reaching far beyond computer science. Historically, naïve approaches to the field have resulted in surprising findings. The development of speech recognition implied an easier way to interact with computers, but this technology was completely rejected by the military and even disheartened them when it was used \cite{Forbus_2004_sketch_maps}. Further research recognized that a deeper understanding of the users and culture of the organization was required despite the belief that voice commands would improve the interface. When designing multimodal interfaces, researchers found that cognitive science was able to dispel many misconceptions and instead provide contrary concrete empirical evidence to help in the design process. These examples highlight the necessity for the field of HCI to reach beyond the field of computer science. Currently, interdisciplinary projects and cross-disciplinary teams are well recognized as research requirements when applicable. This reveals a new question: \textit{What is HCI?} As the lines between disciplines blur, how does HCI fit? And where is the field going? With recent advances in LLMs and related technologies, our technological landscape is rapidly changing requiring HCI to change with it. HCI has stood the test of time, but now more than ever is the opportunity to look back on our understanding and ask ourselves what is next.
\subsection{Human Participation in HCI}

In HCI work, what the ``human'' is and its importance has varied across paradigms. Early HCI focused on the human as {\it user} through a user-centered design approach, where the human was limited to who was processing the information on a screen~\cite{bannonReimaginingHCIMore2011}. Nowadays, the perspective has shifted to be ``human-centered'', where understanding people and their concerns or values is at the forefront. Approaches aligned with this include Norman's Human-Centered Design (HCD) cycle of {\it Observing, Ideating, Prototyping,} and {\it Testing}~\cite{normanDesignEverydayThings2013}, and Friedman et al.'s Value-Sensitive Design method of {\it Conceptual, Empirical,} and {\it Technical} investigations~\cite{friedmanValueSensitiveDesign2013}.
In this so-called ``third wave of HCI'', where technology is ubiquitous and pervasive (existing at a cultural level), humans are full participants in the design process~\cite{bodkerWhenSecondWave2006}, where more humanistic approaches are being integrated into HCI work~\cite{bannonReimaginingHCIMore2011}. 

Beyond HCD and originating in Scandinavia, the participatory design (PD) movement (originally {\it cooperative design}) views the human as a stakeholder through an emphasis on collaboration or co-design among the end-users of a tool and the technology designers~\cite{qiParticipatoryDesignHumanComputer2025}. In HCI, PD studies are applied to a range of application areas, from healthcare and well-being technologies to education and learning ones, in a variety of research phases from exploration to the design of technology~\cite{qiParticipatoryDesignHumanComputer2025}. Going further, approaches hailing from the social sciences such as Action Research~\cite{hayesKnowingDoingAction2014} (also called Participatory Action or Community-Collaborative Research) have been adapted for HCI work~\cite{cooperSystematicReviewThematic2022, tangResearchRelationDocumenting2025}, being proposed as a more equitable way to involve communities in HCI research activities beyond short-term participation. 

As such, human participation in HCI has existed since its inception, and strategies for the inclusion of humans has ranged from a more psychological ``human-as-user'' approach to a more sociological view of humans (and human culture) as active participants and collaborators in research. Recognizing this role, this workshop grounds itself in the multiplicity of human participation in current HCI work.
    
\subsection{Humans in the Loop? The Present and Future of HCI}
The rise of GenAI has transformed the landscape of HCI research. As AI systems increasingly mediate how people make decisions, solve problems, and communicate, researchers have shifted their focus toward understanding how humans collaborate with, interpret, and maintain control over intelligent systems \cite{amershiGuidelinesHumanAIInteraction2019, xuHumancenteredAIPerspective2019}. While early work on automation emphasized the careful balance between machine autonomy and human oversight \cite{parasuramanModelTypesLevels2000}, modern AI systems are moving beyond their role as simple tools and into the role of collaborative partners \cite{xuApplyingHCAIDeveloping2023}. This shift has given rise to the field of Human-AI Interaction (HAI), and with it, a pressing re-examination of what it means for humans to be meaningfully in the loop.

Recent HAI research illustrates both the promise and complexity of this transition. For example, a growing body of work explores Human–AI collaboration in educational settings, where large language models and conversational agents are increasingly used as tutoring systems or learning companions. These systems aim to support learners through personalized feedback and interactive explanations, potentially enabling more adaptive and scalable forms of education \cite{kasneciChatGPTGoodOpportunities2023,wangLargeLanguageModels2024}. There is also HAI work that explores how humans develop trust and reliance working alongside AI systems in complex workflows, underscoring the importance of transparency, explanations, and user control in effective collaboration \cite{ashktorabTrustRelianceEvolving2024}. Despite these benefits, Human–AI collaboration also introduces important risks. Emerging research suggests that as users develop stronger trust and reliance on AI systems, they may become overconfident in the system’s outputs or misunderstand its capabilities \cite{volpatoGenerativeConfidantsHow2026,boInvisibleSaboteursSycophantic2026a}.

These developments highlight a fundamental tension for the future of HCI. While advances in AI promise increasingly autonomous systems, human-centered perspectives argue that maintaining meaningful human involvement is essential for accountability, trust, and ethical decision-making \cite{shneidermanHumanCenteredArtificialIntelligence2020}. As AI systems continue to evolve, we must ask: \textit{do interactive systems still need humans in the loop, or will increasingly autonomous AI make human participation optional, or even obsolete?} This is one of the questions this workshop sets out to explore.

\subsection{Aim of the Workshop}

As a result of this rise of GenAI systems and HAI research, we organize a workshop to:
\begin{enumerate}
    \item Understand how HCI is perceived by the research community,
    \item Explore how human participation exists in HCI work, including (but not limited to) participatory and human-in-the-loop research, and;
    \item Engage in futuring exercises to roadmap the future of HCI research within the socio-cultural zeitgeist of GenAI.
\end{enumerate}

We wish to engage a variety of scholars within the HCI, visualization, and graphics community to discuss what the future of human-centered and -in-the-loop research looks like from the perspective of multiple disciplines. Being aware of the current state, threats to, and benefits of HCI research will help the community understand and prepare for the future of research at-large.

%% file: 2-organizers.tex
\section{Organizers}

All of the organizers of this workshop are graduate students in the HCI Lab at the David R. Cheriton School of Computer Science at the University of Waterloo.

\textbf{Sejal Agarwal} is a master's student researching at the intersection of HCI, AI, and education. Her work focuses on addressing inequities and biases when incorporating GenAI in education.

\textbf{Delara Forghani} is a PhD student researching on building interactive systems using AI and extended reality to develop tutorials about manual skills. 

\textbf{Brandon Lit} is a PhD Student, researching usable security systems, privacy concepts, and in-situ end-user interactions during breach events and cyberattacks.

\textbf{Anthony Maocheia-Ricci} is a master's student researching at the intersection of social computing and creativity. His work currently focuses on creating collaborative tools for ideation and discussion among creative artists.

\textbf{Thomas Driscoll} is a master's student conducting work in HCI and human factors. His work currently focuses on understanding, developing, and improving human performance with input devices.

Where all organizers primarily come from a computer science or engineering perspective in HCI, their individual research areas span multiple facets and sub-disciplines of the field. This breadth in experience enables us to facilitate productive conversation with scholars from a variety of backgrounds and perspectives.

%% file: 3-plans.tex
\section{Workshop Plans}

This workshop is designed as a participatory experience that supports both disciplinary and cross-disciplinary co-creation and reflection. Through guided discussions, we will gather diverse perspectives and collectively re-evaluate the role of the “human” in Human-Computer Interaction. The session is divided into four key phases:

\subsection{Phase 1: Grounding and Positionality}
The workshop begins with a presentation introducing the organizers and their positionality. We will outline the premise and scope of the workshop, posing the central question: \textit{Why does the human matter in an increasingly automated world?}

\subsection{Phase 2: Disciplinary Brainstorming}
To understand the current state of the field, participants will first work in homogeneous groups based on their backgrounds (e.g., UX, AR/VR, education, health). Using a sticky note brainstorming method, groups will identify how they integrate the human element into their work. This activity surfaces diverse, and sometimes conflicting, interpretations of HCI. Each group will then share their insights with the full workshop.

\subsection{Phase 3: The Future of HCI}
Afterward, participants will be reshuffled into interdisciplinary groups and asked to discuss the future of HCI, with a focus on how human-centered design and technologies may evolve in the age of AI. As in the previous phase, participants will use sticky notes to brainstorm before sharing their ideas with the larger group.

\subsection{Phase 4: Synthesis and Conclusion}
The workshop will conclude with a brief presentation synthesizing key themes and insights that emerged across discussions.

%% file: 4-length.tex
\section{Length of the Workshop}

The workshop will run for a total of 1.5 hours, with time allocated across the four phases to balance structure and open discussion. The session will begin with a brief grounding presentation, followed by two interactive brainstorming activities and group discussions, and conclude with a synthesis of key insights.

%% file: 5-cfp.tex
\section{Call for Participation}

We are looking for researchers from a variety of research areas to participate and demonstrate how HCI is used—or could potentially be used—in their fields. While there are no restrictions on the research areas, we wish to be able to include researchers from as diverse a range of areas as possible to highlight how HCI can propagate across multiple domains and practices.
